\documentclass[12pt]{article}

\usepackage{epsf}
\usepackage{epsfig}
\usepackage{cite}
\usepackage{amsmath}
\usepackage{graphics}

\begin{document}

\setlength{\textheight}{21.5cm}
\setlength{\oddsidemargin}{0.cm}
\setlength{\evensidemargin}{0.cm}
\setlength{\topmargin}{0.cm}
\setlength{\footskip}{1cm}
\setlength{\arraycolsep}{2pt}

\renewcommand{\thefootnote}{\#\arabic{footnote}}
\setcounter{footnote}{0}

\newcommand{\gtrsim}{ \mathop{}_{\textstyle \sim}^{\textstyle >} }
\newcommand{\lesssim}{ \mathop{}_{\textstyle \sim}^{\textstyle <} }
\newcommand{\rem}[1]{{\bf #1}}
\renewcommand{\thefootnote}{\fnsymbol{footnote}}
\setcounter{footnote}{0}
\def\thefootnote{\fnsymbol{footnote}}

\begin{titlepage}

\bigskip
\bigskip

\hfill {\tt December 2018}

\begin{center}
{\Large \bf Dark Matter Gravity Waves \\at LIGO and LISA}

\bigskip
\bigskip
\vskip .45in

{\bf Paul H. Frampton\footnote{paul.h.frampton@gmail.com}}
 
 \vskip .3in
 
{\it Department of Mathematics and Physics ``Ennio de Giorgi", \\
University of Salento, Lecce, Italy.}

\vskip 1.0in

\end{center}

\begin{abstract}

\noindent
We study the merger rate of dark matter PIMBHs(=Primordial Intermediate
Mass Black Holes). We conclude that the black holes
observed by LIGO in GW150914 and later events were probably not dark matter
PIMBHs but rather the result of gravitational collapse of very massive stars. To
study the PIMBHs by gravitational radiation will require a detector sensitive
to frequencies below 10Hz and otherwise more
sensitive than LIGO. The LISA detector, expected to come online in 2034,
will be useful at frequencies below 1Hz but a second space-based
detector beyond LISA, sensitive up to 10Hz, will be necessary fully to study dark matter.

\end{abstract}

\end{titlepage}

\renewcommand{\thepage}{\arabic{page}}
\setcounter{page}{1}
\renewcommand{\thefootnote}{\#\arabic{footnote}}

\newpage

\noindent
The first direct detection by LIGO \cite{LIGO} of gravitational waves,  
identified as originating from a merger of 
two black holes with masses $\sim36M_{\odot}$ and $\sim29M_{\odot}$
which occurred at a red shift $Z\sim0.09$, was observed
on September  14, 2015 and, after detailed analysis ensured the data were ironclad,
was first publicly announced on February 11, 2016. This represents the most important cosmological
discovery so far in the 21st century, opening up an entirely new window to observe our Universe.
At last, the direct detection of gravity waves has confirmed the indirect detection deduced
from a binary pulsar by Hulse and Taylor\cite {HulseTaylor1,HulseTaylor2,HulseTaylor3,HulseTaylor4}
and totally vindicates
the prediction of gravity waves made one hundred years ago by Einstein\cite{Einstein1}.

\bigskip

\noindent
It has always been a productive research directions to bridge between
different subfields. In the case of the present Letter, we consider a possible
bridge between the detection of gravity waves\cite{LIGO} and a theory of dark matter
\cite{PHFreview,PHF,Chapline,CF}. In one of these dark matter papers \cite{PHFreview}
the underpinning of the idea that DM(= Dark Matter) $\equiv$ 
PIMBHs(=Primordial Intermediate Mass Black Holes) is discussed thoroughly and
compared to the competing theories where the dark matter constituents are 
elementary particles. It is argued (see especially the Discussion section of \cite{PHFreview})
that, based on the entropy of the universe, DM$\equiv$PIMBHs is strongly favoured over
particle theory alternatives. Therefore, in this Letter, we assume it to be correct.

\bigskip

\noindent
As discussed in the first paper\footnote{An important precursor was by Chapline\cite{Chapline} who
assumed PBHs lighter than the Sun.} suggesting DM$\equiv$PIMBHs \cite{PHF}, published
before the LIGO announcement, the optimal experiment for detecting PIMBHs employs {\it microlensing}. This technique dates back to Einstein\cite{Einstein} and
Paczynski\cite{Paczynski}. Its usefulness was amply demonstrated by the MACHO
Collaboration\cite{Alcock} who observed 17 microlensing events which lensed
stars in the Magellanic Clouds and used a telescope at the Mount Stromlo Observatory
in Australia. Their microlensing events included light curves with duration $\tau$ 
ranging from 34 days to 230 days. From Paczynski's formula\cite{Paczynski}

\begin{equation}
\tau =  \left( \frac{M_{MACHO}}{M_{\odot}} \right)^{1/2} (0.2 M_{\odot})
\label{tau}
\end{equation}

\noindent
this range in $\tau$ translates to $0.2 M_{\odot} \lesssim M_{MACHO} \lesssim 10 M_{\odot}$
for the MACHO mass. The MACHO Collaboration found insufficient mass in MACHOs to constitute all the dark matter, actually less than $20\%$ of the required amount.

\bigskip

\noindent
According to the theory of dark matter proposed in \cite{PHFreview,PHF,CF} the remaining
$80\%$ resides in PIMBHs with masses $M_{PIMBH} > 10M_{\odot}$. If we restrict our attention
to masses readily accessible to microlensing within five years, {{\it i.e.} 
$\tau \lesssim 5y$ then the new masses in play are

\begin{equation}
10 M_{\odot} < M_{PIMBH} < 625 M_{\odot}
\label{625}
\end{equation}

\noindent
The mass range in Eq.(\ref{625}) includes the mass values $\sim 36 M_{\odot}$  and
$\sim 29M_{\odot}$ involved in the first black hole merger, dubbed GW150914, discovered by LIGO\cite{LIGO}.
This does not necessarily imply that the theoretical idea $DM \equiv PIMBHs$ has been either
confirmed or refuted by
LIGO. In the present notes, this question will be discussed
in some detail and our preliminary conclusion will be that the LIGO black holes are probably {\it not}
part of the dark matter, but are more likely not primordial, rather the end products from the
gravitational collapse of very massive stars.

\bigskip

\noindent
Let us anticipate the result that the GW150914 black holes are non-primordial and
retrace the cosmological origin of the binary. The merger
itself happened \cite{LIGO} at $Z=0.09$ which corresponds to cosmic time
$t=12.6Gy$ and is $1.2Gy$ before the present $t_0=13.8Gy$. The capture
and inspiral of two black holes can take gigayears,
although we assume this is less that 7.3Gy which means we do
not study cosmic times less than $t(Z=1) = 5.3Gy$. For later use we
note that the average cosmological matter density at $Z=1$, being
proportional to $(1 + Z)^3$, is $8$ times 
that at the present time $Z=0$.

\bigskip

\noindent
To justify the choice of $Z \leq 1$ for the lifetime of the GW150914 binary, we
use Eq. (5.10) in Peters' 1964 paper\cite{Peters}. Inserting $m_1=m_2 = 30M_{\odot}$
and initial separation $a_0 = \eta R_S$ where $R_S = 90$km is the Schwarzschild radius,
we find a lifetime

\begin{equation}
\tau = 8.3 \times 10^{-12} \eta^4 ~~~{\rm years}
\label{lifetime}
\end{equation}
which is below $7.3$Gy as long as $\eta < 1.7 \times 10^5$. But this implies 
the huge initial separation
$a_0 = 1.5 \times 10^7$ km $\simeq 20R_{\odot}$. This makes our $Z \leq 1$ assumption
very robust. This calculation was done for a circular orbit but, as can be seen from Fig. 2
of \cite{Peters}, for an eccentric orbit the lifetime $\tau$ in Eq.(\ref{lifetime})
becomes even shorter.

\bigskip

\noindent
Consider two PIMBHs approaching each other with some impact parameter
and relative velocity $v$ (which will be written in units of $200km/s$). In
approach they produce a time-varying quadrupole and hence gravity wave (GW)
emission. The cross-section was derived by Quinlan and Shapiro\cite{QS}, and
confirmed by Mouri and Taniguchi\cite{MT}, to be given by

\begin{equation}
\sigma = \pi \left( \frac{85 \pi}{3} \right)^{\frac{2}{7}} R_s^2 \left( \frac{v}{c} \right)^{-\frac{18}{7}}
~~ (pc)^2
\label{sigma}
\end{equation}

\noindent 
where $R_s$ is the Schwarzschild radius given by $R_s=3\eta$ km for 
$M_{PIMBH} = \eta M_{\odot}$.
Let us define $M_{30}$ by $M_{PIMBH} = M_{30} (30M_{\odot})$ and, as stated above,
$v \equiv v_{PIMBH}/(200km/s)$. With these definitions, we find from Eq.(\ref{sigma})
that

\begin{equation}
\sigma \simeq 1.4 \times 10^{-14} M_{30}^2 (v)^{-\frac{18}{7}} ~~ (pc)^2
\label{sigma2}
\end{equation}

\noindent
Using Eq.(\ref{sigma2}), we can calculate the rate of mergers per unit volume
in the dark halo. Let us write the halo mass as $M_{Halo} = M_{12} (10^{12}M_{\odot})$
which defines $M_{12}$. We also define $\rho_{0.002}$ by assuming a uniform dark
matter density $\rho = \rho_{0.002} (0.002 M_{\odot}) /(pc)^3$. The halo volume $V$ is

\begin{equation}
V = \left( \frac{M}{\rho} \right) \simeq (5 \times 10^{14}) 
\left( \frac{M_{12}}{\rho_{.002}}\right) ~~ (pc)^3
\label{V}
\end{equation}

\noindent
The rate of mergers in the halo is

\begin{eqnarray}
N & = & \frac{1}{2} V \left( \frac{\rho}{M_{PIMBH}} \right)^2 \sigma v \nonumber \\
& = & 1.6 \times 10^{-12} \left[ (v)^{-\frac{11}{7}} \rho_{0.002} \right] ~~ yr^{-1}
 \label{N}
 \end{eqnarray}
 
 \noindent
 To estimate the merger rate per unit volume in the visible universe we need
 the mean density $\rho_{DM}$ of dark matter
 
 \begin{equation}
 \rho_{DM} \simeq 3.6 \times 10^{10} M_{\odot} / (Mpc)^3 \equiv 2.3 \times 10^{-29} g / (cm)^3
 \label{rhoDM}
 \end{equation}
 
 \noindent
 and the density $n$ of halos which is
 
 \begin{equation}
 n = \left( \frac{\rho_{DM}}{10^{12} M_{\odot}} \right) \simeq 0.036 (Mpc)^{-3}.
 \label{n}
 \end{equation}
 
 \bigskip
 
 \noindent
 The mean rate of PIMBH mergers $\Gamma$ in the visible universe is
 \begin{equation}
 \Gamma = nN \simeq 2.4\times 10^{-4} \left[v^{-\frac{11}{7}} \rho \right]~~ /(Gpc)^3/yr
 \label{Gamma0}
 \end{equation}  
 
 \bigskip
 
 \noindent
 Taking red shift $Z=0$ values $v\simeq \rho \simeq 1$ the merger rate
 is estimated as
 \begin{equation}
 \Gamma_{theory} \simeq nN \simeq 2.4\times 10^{-4} ~~ /(Gpc)^3/yr
 \label{Gamma1}
 \end{equation}  
 
 \noindent
 which is to be compared with the LIGO estimate\cite{LIGO}
 
 \begin{equation}
 2 ~~ /(Gpc)^3/yr ~~ < \Gamma_{expt} ~~ < 53 ~~ /(Gpc)^3/yr
 \label{LIGOrate}
 \end{equation}
 
 \noindent
 There is significant disagreement between Eq.(\ref{Gamma1}) and Eq.(\ref{LIGOrate}). 
 However, the first LIGO event occurred not
 at $Z=0$ but at $Z\simeq0.09$ \cite{LIGO}. As already mentioned, the beginning
 of the binary capture could not be earlier than $Z=1$. This provides an enhancement
 due to the factor in square brackets in Eq.(\ref{Gamma0}) where, although
 the velocity $v$ does not change significantly, the density $\rho$ is increased
 by a factor $(1+Z)^3 \lesssim 8$ which changes the estimate in Eq.(\ref{Gamma1})
 to
 
 \begin{equation}
 \Gamma_{theory}  \lesssim 1.9 \times 10^{-3} ~~ /(Gpc)^3/yr
 \label{Gamma2}
 \end{equation}  
 
 \noindent
Nevertheless, the marked disagreement between
Eq.(\ref{LIGOrate}) and Eq.(\ref{Gamma2}) leads us to the conclusion that the black holes
involved in the merger which generated the gravity waves observed by LIGO
in \cite{LIGO} are probably not part of the dark matter but are more likely to be non-primordial
and result from gravitational collapse of very massive stars\footnote{We note that in
\cite{Riess} which assumed the start of the merger process was at a larger redshift an enhancement
in the factor $[\left[ (v)^{-\frac{11}{7}} \rho_{0.002} \right]$ by $6\times 10^6$
led to a conclusion that "the possibility that LIGO saw dark matter cannot be
immediately excluded".}.

 \bigskip

\noindent
An interesting question is can LIGO or any of its upgrades detect the dark matter
predicted by the $DM \equiv PIMBHs$ theory? 

\bigskip

\noindent
Two observations can be made immediately:
\noindent
(i) The merger rate calculated in Eq.(\ref{N}) is independent of the PIMBH mass
so increasing this mass has no significant effect on the rate.
\noindent
(ii) Increasing the PIMBH mass does effect the
frequency of the gravity waves as the frequency is inversely proportional to
the mass. The frequencies involved in the LIGO chirp ranged
from 35Hz to 250Hz. The lowest frequency\footnote{This lowest-frequency 
sensitivity limit is principally a result of microseismicity.} to which LIGO is sensitive
is \cite{LIGO2} 10Hz.

\bigskip

\noindent
By scaling the chirp frequencies from those of GW150914, we can check that
when the black hole mass exceeds $105M_{\odot}$ the low frequency beginning
of the chirp is undetectable by LIGO. For black hole mass $M > 750M_{\odot}$, 
LIGO misses the chirp entirely. For the detection of gravity waves from dark matter,
we may thus regard LIGO as a high (audible) frequency detector.

\bigskip

\noindent
There is a second gravity wave detector, LISA(=Laser Interferometer Space
Antenna) \cite{LISA1,LISA2} which will be space-based and expected to begin taking data
in $2034$. Its frequency $f$ sensitivity will be
$0.1mHz < f < 1Hz$ so that its frequency ceiling is significantly below
LIGO's frequency floor. This is unfortunate for the merger of
black holes with $750 M_{\odot}$, for example, where the chirp frequency
ranges from 1.4 Hz to 10Hz.

\bigskip

\noindent
The LIGO and LISA detectors of gravitational radiation were designed
well before the origin and nature of dark matter were understood at  
the level of \cite{PHFreview}. It is clear that at least one additional
gravity-wave detector with greater reach
than LIGO, and sensitive to the crucial frequency range
$1 Hz < f < 10Hz$, will be necessary to investigate the properties
of dark matter. It seems that this additional detector must, like LISA, be
space-based because any ground-based detector,
even underground, cannot be sufficiently sensitive at such low frequencies.

\newpage

\vspace{3.0in}

\begin{center}

\section*{Acknowledgement}

\end{center}

\noindent
We thank Steven Balbus and Selma E. de Mink for useful discussions.


\newpage

\begin{thebibliography}{99}
\bibitem{LIGO}
B.P. Abbott, {\it et al.} (LIGO and Virgo Collaborations),\\
{\it Observation of Gravitational Waves from a Binary Black Hole Merger}.\\
Phys. Rev. Lett. {\bf 116,} 061102 (2016).  \\
{\tt arXiv:1602.03837[gr-qc]}.
\bibitem{HulseTaylor1}
R.A. Hulse and J.H. Taylor, \\
{\it Discovery of a Pulsar in a Binary System}.\\
Astrophys. J. {\bf 195,} L51 (1975).
\bibitem{HulseTaylor2}
J.H. Taylor and J.M. Weisberg,\\
{\it A New Test of General Relativity - Gravitational Radiation \\and the Binary Pulsar
PSR1913+16}.\\
Astrophys. J. {\bf 253,} 908 (1982).
\bibitem{HulseTaylor3}
J.H. Taylor and J.M. Weisberg, \\
{\it General Relativity Geodetic Spin Precession in \\the Binary Pulsar PSR1913+16}.\\
Astrophys. J. {\bf 576,} 942 (2002).
\bibitem{HulseTaylor4}
J.M. Weisberg and Y. Huang,\\
{\it  Relativistic Measurements from Timing the Binary Pulsar \\PSR 1913+16}.\\
Astrophys. J. {\bf 829,} 55 (2016).\\
{\tt arXiv:1606.02744[astro-ph.HE]}.
\bibitem{Einstein1}
A. Einstein,\\
{\it \"Uber Gravitationswellen}.\\
Sitzungber. K. Preuss. Akad. Wiss. 154 (1918).
\bibitem{PHFreview}
P.H. Frampton,\\
{\it On the Origin and Nature of Dark Matter}.\\
Int.J.Mod.Phys. {\bf A33,} 1830030 (2018).\\
{\tt arXiv:1804.03516[physics.gen-ph]}.
\bibitem{PHF}
P.H. Frampton,\\
{\it Searching for Dark Matter Constituents with Many Times\\the Solar Mass.}\\
Mod. Phys. Lett. {\bf A31,} 1650093 (2016).\\
{\tt arXiv:1510.00400[hep-ph]}.
\bibitem{Chapline}
G.F. Chapline,\\
{\it Cosmological Effects of Primordial Black Holes.}\\
Nature {\bf 253,} 251 (1975).
\bibitem{CF}
G.F. Chapline and P.H. Frampton,\\
{\it  Intermediate Mass MACHOs:\\
A New Direction for Dark Matter Searches.}\\
JCAP {\bf 11,} 042 (2016).  {\tt arXiv:1608.04297[gr-qc]}.
\bibitem{Einstein}
A.Einstein,\\
{\it Lens-Like Action of a Star by the Deviation of Light \\
in the Gravitational Field}.\\
Science {\bf 84,} 506 (1936).
\bibitem{Paczynski}
B.Paczynski,\\
{\it Gravitational Microlensing by the Galactic Halo}.\\
Astrophys. J. {\bf 304,} 1 (1986).
\bibitem{Alcock}
C. Alcock, {\it et al.} (The MACHO Collaboration),\\
{\it The MACHO Project: Microlensing Results from 5.7 Years \\
of LMC Observations}.\\
Astrophys. J. {\bf 542,} 281 (2000).  {\tt arXiv:astro-ph/0001272}.
\bibitem{Peters}
P.C. Peters,\\
{\it Gravitational Radiation and the Motion of Two Point Masses.}\\
Phys. Rev. {\bf 136,}  B1224 (1964).
\bibitem{QS}
G.D. Quinlan and S.L. Shapiro,\\
{\it Dynamical Evolution of Dense Clusters of Compact Stars}.\\
Astrophys. J. {\bf 343,} 725 (1989).
\bibitem{MT}
H. Mouri and Y. Taniguchi,\\
{\it Runaway Merging of Black Holes:\\
Analytical Constraint on the Timescale. }\\
Astrophys. J. {\bf 566,} L17 (2002).  {\tt arXiv:astro-ph/0201102}.
\bibitem{Riess}
A.G. Riess, {\it et al.}\\
{\it Did LIGO Detect Dark Matter?}\\
Phys. Rev. Lett. {\bf 116,}. 201301 (2016).\\
{\tt arXiv:1603.00464[astro-ph.CO]}.
\bibitem{LIGO2}
https://dcc.ligo.org/public/0002/T0900288/003/AdvLIGO
\bibitem{LISA1}
T. Robson, N. Cornish and C. Liu,\\
{\it The Construction and Use of LISA Sensitivity Curves}.\\
{\tt arXiv:1803.01944[astro-ph.HE]}.
\bibitem{LISA2}
F. K\"uhnel, A. Matas, G.D. Starkman and K. Freese,\\
{\it Waves from the Centre: Probing PBH and Other Macroscopic Dark
Matter with LISA}.
{\tt arXiv:1811.06387[gr-qc]}.
\end{thebibliography}
\end{document}